\documentstyle[12pt]{article}
\begin{document}

\title{A RENORMALIZATION GROUP STUDY OF HELIMAGNETS IN $D=2+\epsilon$
DIMENSIONS}

\author{P. Azaria\thanks{LPTL, Universit\'e Pierre et Marie Curie,
Tour 16, 4 Place Jussieu, 75230 Paris Cedex 05, France.},\quad
B. Delamotte\thanks{LPTHE Universit\'e Paris 7,
Tour 24 - 5 \'etage, 2 Place Jussieu 75251
Paris Cedex 05, France.},\quad
F. Delduc\thanks{ENSLAPP, ENS Lyon,  46, Allee d'Italie
69364 Lyon cedex 07, France},\quad
and T. Jolicoeur\thanks{
SPhT, CE Saclay, 91191 Gif-sur-Yvette Cedex, France.}}
\maketitle

\begin{abstract}
The non linear sigma model $O(N)\otimes O(2)/O(N-2)\otimes O(2)$ describing
the phase transition of N-components helimagnets is built and studied
up to two loop order in $D=2+\epsilon$ dimensions. It is
shown that a stable fixed point exists as soon as $N$ is
greater than 3 (or equal)
in the neighborhood of two dimensions.
The critical exponents $\nu$ and $\eta$  are obtained.
In the $N=3$ case, the
symmetry of the system is dynamically enlarged at the fixed point
from $O(3)\otimes O(2)/O(2)$ to  $O(3)\otimes O(3)/O(3)\sim O(4)/O(3)$.
We show that the order parameter for Heisenberg helimagnets involves
a tensor representation of $O(4)$ and we
verify it explicitly at one loop order on the value of the exponents.
We show that for large $N$ and in the neighborhood
of two dimensions this nonlinear sigma model describes
the same critical theory as the Landau-Ginzburg linear theory.
As a consequence, the critical behaviour evolves smoothly
between $D=2$ and $D=4$ dimensions in this limit.
However taking into account the old results from the $D=4-\epsilon$
expansion of the linear theory, we show that most likely
the nature of the transition must change between $D=2$ and $D=4$
dimensions for small enough $N$ (including $N=3$).
The simplest possibility is that there exists a dividing
line $N_c (D)$ in the plane $(N, D)$ separating a first-order
region containing the Heisenberg point at $D=4$ and a second-order
region containing the whole $D=2$ axis.
We conclude
 that the phase transition of Heisenberg helimagnets in dimension 3
is either first order or second order with $O(4)$ exponents involving a
tensor representation or tricritical with mean field exponents.
\end{abstract}
\vskip 1.0cm
\centerline{submitted to: {\it Nuclear Physics} B [FS]}
\vskip 1.5cm

SPhT/93-044

PAR/91-27

\newpage
\section{Introduction}
In the recent past, the critical behavior of frustrated spin systems has
been the subject of intensive theoretical, numerical and experimental
studies (see \cite{plumer}\cite{aza1} and references therein
and \cite{aza2}). However,
there is still no definite conclusion about the nature of the phase
transition that occurs in these systems. One of the most striking
features of frustrated models is their non trivial ground state which for
continuous spin models is in general a canted ground state. Well known examples
are incommensurate helimagnets and triangular antiferromagnets among
others. As a consequence of the non collinear ordering the $O(3)$ spin
rotation  group is completely broken in the low temperature phase so that the
relevant order parameter is a rotation matrix instead of a vector as
in ferromagnetic-like models. One may thus wonder if canted spin models belong
to a new universality class. Up to now, no definite answer is known.
Experiments on rare earth helimagnets such as Ho, Tb, or Dy for
example, do not show any clear evidence for a universal critical behaviour.

{}From a theoretical point of view, early renormalization group $(RG)$
studies by Garel and Pfeuty\cite{garel}
found no stable fixed point in the neighborhood of $D=4$
by studying the Landau-Ginzburg theory of a commensurate helimagnet.
This is an example of a fluctuation-induced first order transition. However,
early Monte-Carlo studies
of several canted models \cite{diep}\cite{kawa} found evidence for a
continuous transition in three dimensions. If taken seriously these
numerical results are in contradiction with the $4-\epsilon $ prediction.
Of course, it is notoriously difficult to discriminate between first
and second order phase transitions in Monte-Carlo studies
(the two-dimensional five-states Potts model is a well-known weird
case for example). Strictly speaking, one cannot
exclude the existence of a stable fixed point which manifests itself at
{\sl a finite distance from $D=4$}, unreachable in an $\epsilon$ expansion
around $D=4$. If this happens to be true, then standard perturbative methods
are of no use to study the critical behavior of canted spin models.

There is an alternative perturbative approach to this problem which is the
low temperature expansion of the non-linear sigma (NL$\sigma)$ model suited to
the symmetry breaking scheme of these canted models.
In this paper we focus on a simple commensurate helimagnet which is
the triangular antiferromagnet with N-component classical spins. By
stacking triangular planes, this magnet exists in all integer dimensions
($D\ge 2$).

In the case of the triangular antiferromagnet (AFT) with Heisenberg classical
spins i.e. $N=3$, the massless modes live in the homogeneous space $G/H=
O(3)\otimes O(2)/O(2)$. Some  results from the
$D=2+\epsilon$ expansion of a NL$\sigma$ model based on this coset $G/H$
have been recently reported\cite{aza1}\cite{aza2}. It has been found
that up to two loop a stable fixed point which is the $N=4$ Wilson-Fisher
fixed point shows up in the vicinity of $D=2$. Thus {\sl no new
universality class is required in the case of canted  spin models}.
One meets the general phenomenon of increased symmetry at a critical
point since at this point the model is $O(3)\otimes O(3)= O(4)$ instead of
 $O(3)\otimes O(2)$  symmetric.

In this paper, we extend our previous analysis to the case with $N\ge 3$
components. We build up the relevant nonlinear sigma model
and analyze its RG properties by standard field-theoretic techniques.

If one believes that both the $\epsilon =4-D$ and $\epsilon =D-2$
perturbative results can be extented to non-zero $\epsilon$, in the
neighborhood of $D=2$ and $D=4$, as it is the case for the $O(N)$ models,
 the simplest hypothesis which agrees with both
$\epsilon$ expansions is the following: there is a tricritical surface
separating the basin of attraction of the $O(4)$ fixed point found near
$D=2$ from a first
order runaway region found in the vicinity of $D=4$. The phase transition
of canted magnets is thus
either first order or second order with $O(4)$ or tricritical exponents
(i.e. mean-field in $D=3$). This hypothesis has been previously proposed in
ref.\cite{aza1}. Recent extensive Monte-Carlo studies \cite{bata} performed
directly in D=3 point towards a second-order transition with O(4)
exponents,  a fact that was missed by previous lower-statistics studies.
This means that presumably the critical surface for $N=3$ lies between
D=3 and D=4.

The manifold $G/H$ is topologically equivalent
to $O(3)$ but as metric spaces they are different. The RG properties
of the corresponding non-linear sigma model
are {\it a priori} sensitive to the metric properties.
However the study of purely topological properties can be performed
directly on $O(3)$ as in ref.\cite{kawa2}. The study of defects reveals
the presence of $Z_2$ vortices that are probably liberated in the
high-temperature phase of the strictly two-dimensional AFT model.
In this work we will ignore global aspects and concentrate on
configurations with zero vorticity, leaving for the future the study
of the defects on the phase transition.

In this paper, we present the detailed renormalization group
study of canted spin systems in $D= 2+\epsilon$. In section I we
show how the effective continuum action is obtained from a lattice Hamiltonian
with Heisenberg spins. In section II the group theoretical
construction of the non-linear sigma (NL$\sigma)$ model is presented.
In section III the two loop recursion relations as well as the
Callan-Symanzik $\gamma$-function are given. The critical exponents $\nu$
and $\eta$ are calculated. Special attention is given to the nature
of the order parameter which is shown  to belong to the tensor representation
of $O(4)$. In section IV known results from both
$4-\epsilon$ and $1/N$ expansions are recalled for convenience.
They are discussed and compared with the
$2+\epsilon$ results in section V. Our conclusions are contained in section
VI.

\section{ Continuum limit and effective action}
\subsection{General analysis}
The effective action  that describes the long distance behavior of a
lattice model is obtained by taking the continuum limit of the microscopic
Hamiltonian:
\begin{equation}
H=-\sum_{ij}J_{ij}{\bf S}_i.{\bf S}_j .
\label{ham}
\end{equation}
In this equation the vectors ${\bf S}_i$ are classical Heisenberg spins
with fixed unit length.
 In a ferromagnetic system, this continuum limit is achieved
by letting the spins ${\bf S}$  fluctuate around their common
expectation value. Relative fluctuations between neighboring spins are
assumed to be smooth
enough so that we may replace ${\bf S}_i.{\bf S}_j$ by $(\nabla
 {\bf S}(x))^2$.

When the interaction distribution $\{J_{ij}\}$ leads to a canted ground state
the continuum limit is less obvious since neighboring spins do not
fluctuate around the same mean expectation value. To overcome this
difficulty, one has
to consider the magnetic cell with $n$ sublattices
(${\bf S}^1,\dots, {\bf S}^n$) as the basis of a new
superlattice where the continuum limit is taken. Practically, this procedure
depends on the detailed microscopic
model: lattice symmetry, ground state structure and
interaction parameters. We shall however give qualitative arguments valid for
many canted models.

\noindent Let us define in each elementary cell an orthonormal basis $\{{\bf
e}_a(x)\}$:
\begin{equation}
{\bf e}_a(x).{\bf e}_b(x) =\delta_{ab}\ \ ;\ \  a=1,2,3 \quad ,
\label{eaeb}
\end{equation}
 where $x$ is  a superlattice
index.
We may parametrize our $n$ sublattice spins   ${\bf S}^\alpha(x),
\alpha=1,..,n$ as:
\begin{equation}
 {\bf S}^\alpha(x)=\sum_a C^\alpha_a(x){\bf e}_a(x) .
\end{equation}
In the ground state,  all the $ {\bf S}^\alpha$ are in general
{\it not}
independent. There is in fact a maximum of three of them  which
are independent. Equivalently, there is a minimum of $n-3$ linear combinations
of the ${\bf S}^\alpha (x)$ which have zero expectation value in the ground
state. Such combinations  cannot be part of an order parameter. They
correspond to relative motions of the spins within each unit cell. They are
massive modes with short range correlations and are
thus irrelevant to the critical behavior. We
ignore them by imposing the constraints that {\it locally}, i.e.  within each
unit cell, the spins are  in the
ground state configuration. We call this requirement ``local rigidity''.
Thus, up to an appropriate field redefinition, the order parameter of canted
magnets will be the orthonormal basis $\{{\bf e}_a (x)\}$ defined on each
site of the superlattice.
As a consequence canted magnets are equivalent in the critical
region to a system of interacting solid  rigid bodies. The continuum
effective action ${\cal S}_1$ may be obtained through the standard gradient
expansion of the ${\bf e}_a(x)$ as in ferromagnets:
\begin{equation}
\begin{array}{l}
{\displaystyle{{\cal S}_1={1\over2}\int d^Dx\ \sum\limits_{a=1}^3
p_a\left(\nabla {\bf e}_a\right)^2 ,}}\\
\\
{\bf e}_a(x).{\bf e}_b(x) =\delta_{ab} ,
\end{array}
\label{actionea}
\end{equation}
where the ground state is given by the minimization equations:
\begin{equation}
\begin{array}{l}
\nabla{\bf e}_a(x)=0\nonumber\\
{\bf e}_a(x)= {\bf e}_a^0 .
\end{array}
\label{gs}
\end{equation}
The $p_a, a=1,2,3$ are coupling constants which depend on the
particular lattice model we started with. The partition function $Z$ is :
\begin{equation}
Z=\int D{\bf e}_{1,2,3}(x) \left(\prod_{ab}\delta({\bf e}_a(x).{\bf e}_b(x)
-\delta_{ab})\right)e^{-{\cal S}_1/T} .
\end{equation}
When two couplings $p_a$ vanish, one recovers, integrating over the
corresponding ${\bf e}_a$, the action of the standard non linear sigma model
$O(3)/O(2)$ corresponding to collinear ferro- or anti- ferromagnets.
 In all the other cases, among the nine fields $e_a^i(x)$,
taking into account the constraints (\ref{eaeb}),
 one sees that there are three independent fluctuating fields corresponding
to the three Goldstone modes, or spin waves, resulting from the breakdown
of the $O(3)$ group. Each one corresponds to infinitesimal rotations around
each of the ${\bf e}_a(x)$'s. The couplings $p_a$ are the associated stiffness
constants which depend on the detailed microscopic model. They are deeply
connected to the symmetry properties of the lattice Hamiltonian  as we shall
see.

In addition to the usual $O(3)$ rotational invariance, the symmetry
group $G$ of the Hamiltonian  contains, in general,  a discrete group
${\cal T}$ of
transformations \{$T^s$\}  mixing together the sublattices ${\bf S}^\alpha$, or
equivalently the
${\bf e}_a$. These transformations may belong to the space group of
the lattice, as in triangular antiferromagnets, but may be some more
complicated objects, such as ``gauge" transformations as in the
Villain lattice\cite{yosefin}.
The order parameter $\{{\bf e}_a\}, a=1,2,3$ thus transforms under
$G$ as:
\begin{equation}
\begin{array}{l}
{\displaystyle{{ e}_a^i\to \sum_jU_{ij}\ { e}_a^j\ ;\ U\in O(3),}}\\
\\
{\displaystyle{{\bf e}_a\to \sum_b(T^s)_{ab}\ {\bf e}_b\ ;\ T^s\in {\cal T}}}.
\end{array}
\end{equation}
The requirement that the
action ${\cal S}_1$ should be
 invariant under the group ${\cal T}$ implies several
 relations between the $p_a$'s. In general, the ${\bf e}_a$'s span
reducible representations of the group
${\cal T}$. Depending on the number of these representations, some of the
 coupling constants $p_a$ may be equal. If there are
three irreducible representations of dimension 1, all the $p_a$ are different.
If there is one representation of dimension 2 and one of dimension 1, as it
is the case in the triangular  lattice where ${\cal T}$ is $C_{3v}$, two
coupling constants are equal: $p_1=p_2$. In this case, since the
action is quadratic in the fields,
 the invariance under the discrete group ${\cal T}$ is enlarged to a
continuous invariance group $O(2)$ generated by:
\begin{equation}
\left(
{\bf e}_1, {\bf e}_2, {\bf e}_3
\right)
\to
\left(
{\bf e}_1, {\bf e}_2, {\bf e}_3
\right)
\left(
\begin{array}{ccc}
\cos\theta &\sin\theta &0\\
-\sin\theta & \cos\theta & 0\\
0 & 0 & 1
\end{array}
\right),
\label{tr1}
\end{equation}
and
\begin{equation}
\left(
{\bf e}_1, {\bf e}_2, {\bf e}_3
\right)
\to
\left(
{\bf e}_1, {\bf e}_2, {\bf e}_3
\right)
\left(
\begin{array}{ccc}
0 &1 &0\\
1& 0 & 0\\
0 & 0 & 1
\end{array}
\right).
\label{tr2}
\end{equation}
 The action ${\cal S}_1$ is thus  $O(3)\otimes O(2)$ invariant in this case.

Finally, if there is only one representation
 of dimension 3, as it is the case in lattices with tetragonal symmetry,
$p_1=p_2=p_3$ and ${\cal S}_1$ is invariant under $G=O(3)\otimes O(3)$.
To summarize,  depending on the  symmetry
group of the lattice, ${\cal S}_1$ can be symmetric under $O(3)\otimes O(p)$
with $p=1,2$ or 3.

Since any rotation matrix $R\in$ SO(3)  is given by an orthonormal set
of three vectors, we can gather the  ${\bf e}_a$'s into a rotation matrix $R$:
\begin{equation}
R(x)=\big({\bf e}_1(x){\bf e}_2(x){\bf e}_3(x)\big),
\end{equation}
and the action ${\cal S}_1$ can be written into a different, but
{\it equivalent}, form:
\begin{equation}
{\cal S}_2={1\over2}\int d^Dx\  Tr\left(P(\partial R^{-1})(\partial R)\right),
\label{actionp}
\end{equation}
where $P$ is the diagonal matrix: $P=diag (p_1,p_2,p_3)$ and $R\in$ SO(3).

 Using $R$, the symmetry operations on the ${\bf e}_a$ can be written in a
compact form.
The action ${\cal S}_2$ is invariant under left $O(3)$
transformations $R\to
UR\ ,\ U\in$ O(3) and right transformations: $R\to RV$ where $V$ belongs to
the $O(p)$ group which commutes with matrix $P$. We
find again  that depending on
the value of $p$, ${\cal S}_2$ is invariant under the
group $G=O(3)\otimes O(p); p=1,2,3$. The
right $O(p)$ invariance  reflects the original discrete symmetry of the
microscopic Hamiltonian since it mixes the ${\bf e}_a$
 while the left $O(3)$ is the usual rotational
symmetry.
The  discrete symmetry group of the Hamiltonian acts, in the continuum
limit, as the $O(p)$  group. This is an artefact of the
continuum limit and has no dynamical consequences since the number of
Goldstone modes is given by the breaking of the $O(3)$ spin rotation
group only. Indeed the number $n$ of Goldstone modes   resulting
from the symmetry breaking  $G\to H$, where $H$ is the subgroup of $G$ which
leaves the ground state invariant, is equal to the number of broken
generators of G:
$n=dim\ Lie (G) -dim\ Lie (H)$. In our case, the isotropy subgroup $H$
consists of the transformations of $G$  leaving the orthonormal
basis ${\bf e}_a^0$ invariant or equivalently using Eq.(\ref{gs}) and the
above $U$ and $V$ transformations:
\begin{equation}
H\ :\ R_0\to {\hat{V}}R_0V=R_0,
\end{equation}
where ${\hat{V}}\in O(p)\subset O(3)$ and is determined once $V$ is chosen (see
Eq.(\ref{vchapeau} for an explicit expression of  ${\hat{V}}$ in a
particular example).
This particular subgroup $H$ is called the diagonal group
$O(p)_{diag}$ of
a subgroup $O(p)$ in $O(3)$ times the  $O(p)$  of $G$ acting on
the right. The symmetry breaking patterns described by action
(\ref{actionea})
and equivalently by action (\ref{actionp}) are therefore
$G/H = O(3)\otimes O(p)/O(p)_{diag}$ with $p=1,2,3$ depending
on matrix
$P$. These are all the possible symmetry breaking schemes that can undergo a
frustrated Heisenberg spin model. In any case, there are three Goldstone
modes. Before ending this section, let us emphasize that in the non linear
sigma model, the dynamical properties depend only on the {\it geometry} of
the coset space $G/H$  and not on the field parametrization. Contrary
to the Landau-Ginzburg model, the vanishing dimension of the order parameter
in two dimensions allows infinitely many different parametrizations of the
theory, differing in (possibly non linear) field redefinitions. The choice of
one particular parametrization may be useful in discussing symmetry
properties or renormalization group equations but does not change the
physics. In particular we have seen
that the actions ${\cal S}_1$ and ${\cal S}_2$ are equivalent. There is
another equivalent form of ${\cal S}_1$ and ${\cal S}_2$ that we shall
use for the discussion of the $O(3)\otimes
O(2)/O(2)$ case:
\begin{equation}
{\cal S}_3 ={1\over2}\int d^Dx \left(g_1\left(\nabla
{\bf e}_1^2 + \nabla
{\bf e}_2^2\right)
+ g_2\left({\bf e}_1\nabla{\bf e}_2 - {\bf e}_2\nabla{\bf e}_1\right)^2\right),
\label{actioncourant}
\end{equation}
with:
\begin{equation}
{\bf e}_a.{\bf e}_b =\delta_{ab}\ ,\ g_2=-{1\over2}p_2\ ,\
g_1=p_1+p_2 .
\end{equation}
The latter expression of the action is obtained from (\ref{actionea}) by
integrating over the
constraint ${\bf e}_3={\bf e}_1\wedge {\bf e}_2$ and the relations
among the coupling constants are derived in Appendix B.

\subsection{The particular case of the antiferromagnetic triangular lattice}

We shall consider the antiferromagnetic triangular
lattice with $N$-component spins, $N\ge 3$, as an example. Let us start by the
case $N =3$. The symmetry group of the system is the product of
the rotation group $O(3)$ acting on the spin
components times the space group of the triangular lattice. The Hamiltonian
density must then be $G=O(3)\otimes C_{3v}$ invariant. The three spins
${\bf S}_i$, $i=1,2,3$ of the elementary plaquette are co-planar in the
ground state with the well-known 120 degrees structure. Then, we expect that
only two vectors of the triad $({\bf e}_1, {\bf e}_2,{\bf e}_3)$ are necessary
for the decomposition of the ${\bf S}_i$.

i) $\Sigma = {\bf S}_1 + {\bf S}_2 + {\bf S}_3$ spans the
trivial representation of $C_{3v}$. This linear combination of the spins
has a vanishing expectation value at $T=0$. It cannot be an order
parameter and corresponds to massive modes.

ii) The two vectors:
\begin{equation}
\left(
\begin{array}{l}
{\bf e}_1\\
\\
{\bf e}_2
\end{array}
\right)
\propto
\left(
\begin{array}{l}
-{{\sqrt3} + 1\over 2}{\bf S}_1 +{{\sqrt3} - 1\over 2}{\bf S}_2 + {\bf S}_3\\
\\
{{\sqrt3} - 1\over 2}{\bf S}_1 -{{\sqrt3} + 1\over 2}{\bf S}_2+ {\bf S}_3
\end{array}
\right)
\end{equation}
span the two dimensional representation of $C_{3v}$. They have a non vanishing
expectation value in the low temperature phase and can  thus be taken as
an appropriate order parameter.
The ``local rigidity'' constraint of section (2.1) means here:
\begin{equation}
\Sigma (x) = <\Sigma (x)>= 0.
\end{equation}
This allows fluctuations of the spins between cells but not within the
cells. This constraint is consistent with
the symmetry since $\Sigma$ is a scalar for $C_{3v}$. Once the constraint is
imposed, it is straightforward to show
that ${\bf e}_1(x)$ and ${\bf e}_2(x)$ are orthonormal by use of ${\bf S}_i^2
=1$.
The action for the triangular lattice is then:
\begin{equation}
{\cal S}_1 = {1\over 2}\int d^Dx \left( p_1\left( (\nabla {\bf e}_1(x))^2
+ (\nabla {\bf e}_2(x))^2\right) \right).
\label{jenaimarre}
\end{equation}
Dombre and Read have obtained, from a direct microscopic derivation
\cite{dombre}:
\begin{equation}
p_1 ={{\sqrt{3}}\over4} {J\over T} .
\end{equation}
Note that the original  $C_{3v}$ invariance has been enlarged in
Eq.(\ref{jenaimarre}) to a $O(2)$ group  given by
(\ref{tr1},\ref{tr2}): $G=O(3)\otimes O(2)$.
We will see that this action is not stable under renormalization. The
most general renormalizable action which is compatible with the symmetry
$O(3)\otimes O(2)$ is given by equation (\ref{actioncourant}). Its general
form is stable under renormalization so that we shall work only with
it in the following.

It is easy to generalize the above action to the case where the fields
${\bf e}_a(x)$ have $N> 3$ components. The symmetry group $G$ is in this case
$O(N)\otimes O(2)$. The ground states are given by eq.(\ref{gs}).
Let us choose one of them, for example:
\begin{equation}
({\bf e}_1^{(0)}, {\bf e}_2^{(0)}) =
\left(
\begin{array}{ll}
0&0\\
.&.\\
1&0\\
0&1
\end{array}
\right).
\label{GS}
\end{equation}
The unbroken symmetry group $H$ of the low-temperature phase  is then
the set of matrices leaving this configuration invariant:

\begin{equation}
\left(
    \begin{array}{cc}
    {\large{O}(N-2)}&0\\
    0&\begin{array}{cc}
    \cos\theta&-\sin\theta\\
    \sin\theta&\cos\theta
    \end{array}
    \end{array}\right)
    \left(
    \begin{array}{c}
    0\\
    .\\
    1\\
    0
    \end{array}
    \right.
    \left.
    \begin{array}{c}
    0\\
    .\\
    0\\
    1
    \end{array}
    \right)
    \left(\begin{array}{c}
    \cos \theta\\
    -\sin\theta
    \end{array}
    \right.
    \left.\begin{array}{c}
    \sin\theta\\
    \cos\theta
    \end{array}
    \right)
    =\left(
    \begin{array}{c}
    0\\
    .\\
    1\\
    0
    \end{array}
    \right.
    \left.
    \begin{array}{c}
    0\\
    .\\
    0\\
    1
    \end{array}
    \right).
\label{vchapeau}
    \end{equation}
This group $H$ consists in two subgroups: a $O(N-2)$ and a
diagonal $O(2)$. Therefore the action ${\cal S}_1$ describes the symmetry
breaking pattern $G\to H= O(N)\otimes O(2)\to O(N-2)\otimes O(2)_{diag}$.

\section{Group theoretical construction of the non linear sigma model}

In the last section, we have derived three equivalent forms  of the relevant
action for  Heisenberg frustrated magnets ${\cal S}_1$, ${\cal S}_2$,
${\cal S}_3$ (eq.(\ref{actionea},\ref{actionp},\ref{actioncourant})). Each
of them has its own interests and shortcomings. Action ${\cal S}_1$ is closely
related to the microscopic Hamiltonian while action ${\cal S}_2$ is
suited to the discussion of the symmetry properties. Finally
action ${\cal S}_3$
offers a possible large $N$, $N\ge3$, generalization which we shall
discuss in detail. The particular form  of the action is irrelevant
near $D=2$ since the RG properties depend only on the geometry of the
manifold $G/H$ \cite{friedan}.  These intrinsic
properties will be formulated in the language of group theory which provides
an abstract but powerful framework\cite{friedan}.
We shall first deal with the standard $O(N)/O(N-1)$ model.
The construction of the $ O(N)\otimes O(2)/O(N-2)\otimes O(2)$ model
is presented after this warm-up example.

\subsection{The $O(N)/O(N-1)$ partition function}

 The partition function of the $O(N)/O(N-1)$ model is \cite{polyakov,brezin}:

    \begin{equation}
    Z= \int D{\bf S}\ \delta\left({\bf S}^2(x)-1\right)\
\exp \left({-{1\over 2T}\int d^Dx (\partial { S})^2}\right).
    \label{fonctpart}
    \end{equation}
The functional delta selects the configurations of ${\bf S}(x)$ with
unit length. We can take advantage of this delta to integrate out one degree of
freedom in ${\bf S}(x)$. Let us choose  ${\bf u}$, ${\bf u}^2=1$, collinear to
the magnetization  and write ${\bf S}(x)$ as:
\begin{equation}
{\bf S}(x) = \sigma (x){\bf u} + {\bf\pi}(x) \hspace{0.6cm};
\hspace{0.6cm} {\bf\pi}(x) \bot  {\bf u} \hspace{0.6cm};
\hspace{0.6cm} \sigma^2 + \pi^2 =1 .
\label{sigma}
\end{equation}
After integrating out $\sigma(x)$, $Z$ can be rewritten as:
    \begin{equation}
    Z=
    \int_{\vert \pi\vert\le 1} D{\bf\pi} \ \exp \left({-{1\over 2T}
    \int d^Dx\left((\partial{\bf\pi})^2 +
    (\partial\sqrt{1-{\bf\pi}^2})^2\right)}\right).
    \label{ZON}
    \end{equation}
The low temperature $T$ perturbative calculation of
(\ref{ZON}) starts from small
fluctuations around the ground state: $<{\bf S}>= {\bf u}$. They correspond
to the excitations of the ${\bf\pi}$-field and are the usual spin waves.
The ${\bf\pi}$'s consist of the $N-1$ Goldstone modes coming from the
breaking of $O(N)$
down to the rotation group $O(N-1)$ that leaves the ground state
invariant, i.e. the $O(N-1)$ around the ${\bf u}$-direction. Once the symmetry
breaking pattern $O(N) \to O(N-1)$
is given the NL$\sigma$ model is entirely determined up to the coupling
constant which in this case is the temperature.

We present now a matrix formulation of the  $O(N)/O(N-1)$ model. Let us
choose a ground state  ${\bf S}^0={\bf u}$. We can write the
${\bf S}(x)$ field as:
    \begin{equation}
    {\bf S}(x) = R(x){\bf S}^0 ,
    \label{R(x)}
    \end{equation}
where $R(x)$ is the $O(N)$ matrix sending ${\bf S}^0$ onto ${\bf S}(x)$.
The partition function $Z$ can be rewritten as:

 \begin{equation}
Z= \int D{R}\  \exp \left({-{1\over 2T}\int d^Dx Tr(K(\partial R^{-1}
\partial R))}\right),
\label{cestlafin}
\end{equation}
where
\begin{equation}
K_{\alpha\beta}= S_{\alpha}^OS_{\beta}^O .
\label{k}
\end{equation}
In this last equation the indices are those of vectors
$(\alpha,\beta)=1,..,N$ and $R\in O(N)$.
The relationship between ${\bf S}(x)$ and $R(x)$ is not bi-univoque
since for any rotation matrix $h(x)$ leaving ${\bf S}^0$ invariant:
    \begin{equation}
     h{\bf S}^0={\bf S}^0 ,
    \end{equation}
    one has:
    \begin{equation}
    R(x)h(x){\bf S}^0=R(x){\bf S}^0 ={\bf S}(x) .
    \end{equation}
    As a consequence, the action  is locally (i.e. gauge) right invariant
    under the transformation:
    \begin{equation}
    R^h(x) = R(x)h(x)\ ,\ h\in H=O(N-1) .
    \label{transfojauge}
    \end{equation}
Some degrees of freedom in $R\in O(N)$ are thus unphysical.
To obtain a bi-univoque representation in terms of matrices we have
to choose one unique element in each equivalence class $R^h$, that
is to fix the gauge.
The set of these equivalence classes is the set of $O(N)$ rotations up to a
$O(N-1)$ rotation: it is $O(N)/O(N-1)$. We can easily find one element
per equivalence class in terms of the physical ${\bf\pi}$-field. Let us
write $R(x)$ and $h(x)$ as:
    $$
    R(x)=\left(
    \begin{array}{c}
    \large{A}\\
    ^t{\bf V}'
    \end{array}
    \right.
    \left.
    \begin{array}{c}
    {\bf V}\\
    B
    \end{array}
    \right)
    \qquad
    h(x)=\left(
    \begin{array}{c}
    h'\\
    0
    \end{array}
    \right.
    \left.\begin{array}{c}
    0\\
    1
    \end{array}
    \right)
    \quad
    h'\ \in\  O(N-1) .
    $$
The matrix ${A}$ is $(N-1)\times (N-1)$,  ${\bf V}$ and ${\bf V}'$ are a
$N-1$-component vectors and $B$ is a scalar.
We use relation (\ref{transfojauge}) to eliminate as many degrees of
freedom in $R(x)$ as possible. It is convenient to choose:
    \begin{equation}
    h'(x)=A^{-1}\sqrt{A\ ^tA}.
\label{hash}
    \end{equation}
This leads to exactly one element per class given by:
    \begin{equation}
    L= \left(\begin{array}{c}
    \sqrt{1_{N-1}-{\bf V} \ ^t{\bf V}}\\
    -^t{\bf V}
    \end{array}
    \right.
    \left.
    \begin{array}{c}
    {\bf V}\\
    \sqrt{1-{\bf V}^2}
    \end{array}
    \right).
    \end{equation}
We identify ${\bf V}$ by applying $L$ to ${\bf S}^0$, eq.(\ref{R(x)}):
     \begin{equation}
    \left(\begin{array}{c}
    {\bf\pi}\\
    \sigma
    \end{array}
    \right)=
    L\left(\begin{array}{c}
    0\\
    1
    \end{array}
    \right).
\label{defL}
\end{equation}
The element $L$ can thus be written entirely in terms of the $\pi$ fields:
    \begin{equation}
    L(\pi^i)= \left(\begin{array}{c}
    \sqrt{1_{N-1}-{\bf \pi} \ ^t{\bf\pi}}\\
    -^t{\bf \pi}
    \end{array}
    \right.
    \left.
    \begin{array}{c}
    {\bf \pi}\\
    \sigma
    \end{array}
    \right).
    \end{equation}
The set of $L$-matrices is such that to any
$\pi$ corresponds a unique $L(\pi)\in O(N)/O(N-1) $. The quantity
$(L^{-1}\partial L)$ belongs to the Lie algebra $Lie(G)$ of $G=O(N)$
and we have:
   \begin{equation}
   (L^{-1}\partial L)=(L^{-1}\partial L)_{G-H}+(L^{-1}\partial L)_{H}
   \label{ldl}
   \end{equation}
    where $(L^{-1} \partial L)_{H}$ is in  $Lie(H)$.
The partition function $Z$ can be finally written as:
\begin{equation}
Z= \int D{\pi}\  e^{-{1\over T}S},
\label{zl}
\end{equation}

\begin{equation}
S=-{1\over 2}\int d^Dx\ Tr([( L^{-1} \partial L)_{G-H}]^2).
\label{actiona}
\end{equation}
We have used the fact that $K$ is a projector:  $K( L^{-1} \partial L)_{H}=0$.
The partition function in Eq.(\ref{cestlafin}) is globally $G$-invariant
and locally (i.e.gauge) $H$-invariant. Once
a gauge choice is made (as in (\ref{hash},\ref{defL})) no $H$-transformations
are allowed in (\ref{zl},\ref{actiona})
and the
$G$-transformations are in general not compatible with the gauge choice,
i.e. they do not preserve the form of matrices $L$. This means that a
$G$-transformation must be accompanied by a $H$-gauge-restoring-transformation:
 \begin{equation}
    L(\pi') = gL(\pi) h(g,\pi).
    \label{ltransfo}
    \end{equation}
    Thus, $G$ is non linearly realized on the $\pi$-fields.  This is
    completely different from
    the Landau-Ginzburg model where $G$ is linearly realized on the $\pi_i$
    fields.

Equation (\ref{zl}) is the general expression for the partition function of a
NL$\sigma$ model defined on a coset space $G/H$. This coset space can be
viewed as a metric manifold so that it is convenient to formulate the theory
in the language of differential geometry.

Since $L^{-1}\partial L$ belongs to $Lie(G)$, eq.(\ref{ldl}) can be
rewritten as:
\begin{equation}
L^{-1}\partial_\mu L = e_\mu^I T_I +\omega_\mu^a T_a ,
\label{viel}
\end{equation}
where the $T_a$'s are the generators of $Lie(H)$ while the $T_I$'s are
generators in $Lie(G)-Lie(H)$. $e_\mu^I$ and $\omega_\mu^a$ are respectively
the vielbein and the connection in the tangent space of $G/H$.  Under
(\ref{ltransfo}) they transform as:
\begin{equation}
    \left\{
    \begin{array}{ll}
    {e'}_\mu^IT_I &= h^{-1}(x)(e_\mu^IT_I)h(x),\\
    {\omega'}_\mu^aT_a &= h^{-1}(x)(\omega_\mu^a T_a) h(x) +h^{-1}(x)
\partial_\mu h(x).
    \end{array}
    \right.
    \end{equation}
The $T_I$'s span a representation of $H$ since:

\begin{equation}
[T_a,T_I] = {f_{aI}}^JT_J .
\label{transt}
\end{equation}
As a consequence, the $e_\mu^I$'s span a linear representation of $H$.
Using (\ref{viel}), action $S$ in eq.(\ref{actiona}) can be written as:

 \begin{equation}
S={1\over2}\int d^D x\ e_\mu^I e_\mu^J\eta_{IJ},
\label{toto}
\end{equation}
where $\eta_{IJ}$ is the tangent space metric given by:

\begin{equation}
\eta_{IJ}= -Tr(KT_IT_J),
\end{equation}
with $K$ the projector on $G-H$. In the $O(N)/O(N-1)$ case it is given by
eq.(\ref{k}). In this case, the $N-1$ generators $T_I$'s of
$Lie(O(N))-Lie(O(N-1))$ span the vector
representation of $O(N-1)$ so that there is only one coupling constant:
$\eta_{IJ}= \eta\delta_{IJ}$. However,  in general, $\eta_{IJ}$ is a diagonal
matrix with several different couplings. The number of these couplings is
the number of quadratic invariants under
transformation (\ref{transt}) constructed with the $e_\mu^I$'s.
For a symmetric space there is only one such invariant. This is the case
of $O(N)/O(N-1)$ for example. For a non-symmetric homogeneous
space such as $O(N)\otimes O(2)/O(N-2)\otimes O(2)$ this number is
larger than one.
The formula $e_\mu^I = e_i^I\partial_\mu\pi^i$ leads to the more conventional
form eq.(\ref{ZON}) of the action of the NL$\sigma$ model defined on a coset
 space viewed
as a metric space equipped with the metric $g_{ij}(\pi)=e_i^Ie_j^J\eta_{IJ}$
\cite{friedan}:

\begin{equation}
Z=\int_{\vert \pi\vert\le 1} D{\bf\pi} \
\exp \left( -{{1\over 2T}\int d^Dx\ g_{ij}(\pi)
    \partial\pi^i\partial\pi^j} \right).
\label{toto2}
\end{equation}
 Eq.(\ref{toto}) and eq.(\ref{toto2}) provide alternative  descriptions of
NL$\sigma$ models defined on a coset space $G/H$ in terms of purely local
geometrical quantities of the manifold $G/H$ such as for example the
metric,  Riemann and  Ricci tensors. It is equivalent
to work either on the manifold itself (\ref{toto2}) or in the tangent space
(\ref{toto}). For practical calculations, it is extremely convenient to use
the tangent space formulation we have discussed above. It can be shown that
the geometrical quantities such as the Riemann tensor depend only in
tangent space on the Lie algebras $Lie(G)$ and
$Lie(H)$. More precisely they depend on the structure constants defined by the
following commutation rules:

\begin{equation}
\begin{array}{l}
[T_a,T_I] = {f_{aI}}^JT_J ,\\
{[T_a,T_b] = {f_{ab}}^cT_c},\\
{[T_I,T_J] = {f_{IJ}}^KT_K + {f_{IJ}}^aT_a},
\end{array}
\label{fstructure}
\end{equation}
where $T_a\in Lie(H)$ and $T_I\in Lie(G)-Lie(H)$.

\subsection{The ${O(N)\otimes O(2)/ O(N-2)\otimes O(2)_{diag}}$ partition
function}

 In the case of the  ${O(N)\otimes O(2)/ O(N-2)\otimes O(2)_{diag}}$ model,
 the order parameter is the set of $N$-component vectors
 $({\bf e}_1,{\bf e}_2)$ and the action is given by action ${\cal S}_3$
 eq.(\ref{actioncourant}). Let us define the order parameter
  as the rectangular matrix:

 \begin{equation}
 \Phi =({\bf e}_1,{\bf e}_2).
 \end{equation}
 The  $O(N-2)\otimes O(2)$ transformations can be written:

    \begin{equation}
    ^t\Phi ' = ^tr(x)\ ^t\Phi\ ^tR(x),
\label{transphi}
    \end{equation}
where $R\in O(N-2)$ and $r\in O(2)$. The ground state Eq.(\ref{GS}) is
invariant under the transformations:

    \begin{equation}
    ^t\Phi^0 = h_1(x) ^t\Phi^0 H(x),
    \end{equation}
    with $h_1\in O(2)$ and
    \begin{equation}
    H(x)=
    \left(
    \begin{array}{cc}
    h_2(x) & 0\\
    0 & h_1^{-1}(x)
    \end{array}
    \right),
    \qquad
    h_2\ \in\ O(N-2).
    \end{equation}
    Thus, the matrices $r(x)$ and $R(x)$ are defined up to the following local
    transformations:
    \begin{equation}
    \left\{
    \begin{array}{lll}
    r(x) & \to &r(x)\ h_1(x)\\
    R(x) & \to & R(x)\ H(x)
    \end{array}
    \right.
    \label{transformations}
    \end{equation}
    In the low temperature phase we can rewrite $\Phi$ in terms of the $2N-3$
    Goldstone modes:
    \begin{equation}
    \Phi(x)=
    \left(
    \begin{array}{c}
    \pi(x)\\
    \omega(x)\sqrt{1_2 - {^t\pi\pi}}
    \end{array}
    \right),
    \end{equation}
    where $\pi$ is a $(N-2)\times 2$ matrix and $ \omega(x)\in O(2)$. The
$\pi_i^\alpha$, $i=1,\dots ,N, \alpha=1,2$ transform as two independent
vectors under $O(N-2)$ and as a vector under $O(2)$. $\omega(x)\sqrt{1_2
- {^t\pi\pi}}$ represents one extra degree of freedom which is scalar under
both $O(N-2)$ and $O(2)$.
    One can use the gauge freedom (\ref{transformations}) to go from
    a general element $R(x)\otimes r(x)$ of $O(N)\otimes O(2)$ to the
    unique element in the same gauge orbit $L\otimes {\bf 1}_2$:

    \begin{equation}
    L(\pi(x),  \omega(x)) =
    \left(\begin{array}{cc}
    \sqrt{1-\pi\ ^t\pi} & \pi\\
    - \omega(x) ^t\pi &  \omega(x)\sqrt{1 - {^t\pi\pi}}
    \end{array}
    \right).
    \label{LpiO}
    \end{equation}
    The matrix $L$ thus parametrizes the coset space $O(N)\otimes
O(2)/O(N-2)\otimes O(2)_{diag}$. Note that this matrix is the same as for the
coset space $O(N)/O(N-2)$.

In fact this is not accidental and we will use the following property
to simplify our study:
very generally the coset
spaces $G\otimes X/H\otimes X_{diag}$ (where $X$ is the maximal subgroup
of $G$ commuting with $H$) and $G/H$ are topologically equivalent.
We can thus work directly with the coset $G/H$
keeping in mind that we search for an action which has $G\otimes X$
as symmetry (isometry) group.

The vielbein of $G/H$
defined as in eq.(\ref{viel}) decompose into two irreducible representations
under the action of $H\otimes X$. $X$ itself spans the adjoint
representation of $X$ and is a scalar under $H$. $G-H-X$ is irreducible
because $H\otimes X$ is maximal in $G$, stated otherwise $G/H\otimes X$
is a symmetric space. Thus, the two
projected matrices $(L^{-1}\partial L)_{\vert G-H-X}$ and
$(L^{-1}\partial L)_{\vert X}$ transform independently under the right
action of the $H\otimes X$ group so that there are two independent
couplings $\eta_1$ and $\eta_2$. We are thus led to the action:
\begin{equation}
S=-{1\over 2}\int d^Dx\left( \eta_1tr(L^{-1}\partial L)_{\vert G-H-X}^2
+\eta_2tr(L^{-1}\partial L)_{\vert X}^2 \right).
\label{genac}\end{equation}
Denoting by $I$ the indices of $Lie(G)-Lie(H)$, and among them by $\alpha$
the indices of $Lie(X)$, this action may be rewritten as:
\begin{equation}
S={1\over 2}\int d^Dx\ e^I_\mu e^J_\mu \eta_{IJ}.
\end{equation}
where the tangent space metric $\eta_{IJ}$ is given by:
\begin{equation}
\eta_{IJ}=-\eta_1tr(T^IT^J)-(\eta_2-\eta_1)\delta_{I\alpha}
\delta_{J\beta}tr(T^\alpha T^\beta).
\label{eta}\end{equation}
We recall that in our case $T_I\in Lie(O(N))-Lie(O(N-2))$, $T_\alpha\in
Lie(O(2))$ and
$T_a\in Lie(O(N-2))$ and that the corresponding algebra is given
in (\ref{fstructure}).
 Action (\ref{genac}) is completely equivalent to the action  ${\cal S}_3$
 we have obtained from the continuum limit (\ref{actioncourant}). We prove
it in appendix B and derive the
relations between the couplings $g_1, g_2$ entering in  ${\cal S}_3$
and $\eta_1,\eta_2$: $\eta_1=g_1/2\ ;\ \eta_2= g_1+2g_2$.

\section{Renormalization of the $NL\sigma$ model in
$D=2+\epsilon$}

\subsection{General case}

The renormalizability in $D=2+\epsilon$ of
NL$\sigma$ models defined on coset spaces $G/H$ was studied by D.H.
Friedan\cite{friedan}. The $\beta$ function gives the
 evolution of the metric $g_{ij}(\pi)$
with the scale:
\begin{equation}
{\partial g_{ij}\over \partial l}= \beta_{ij}.
\label{evolution}
\end{equation}
At two loop order it is given by the following expression:

\begin{equation}
\beta _{ij}(g)=   -\epsilon g_{ij} + R_{ij} + {1\over2} T R_{ipqr}R_{jpqr} +
O(T^2).
\label{beta}
\end{equation}
where:
$R_{ij}$ and $R_{jpqr}$ are the Ricci and the Riemann tensors of the manifold
$G/H$ equipped with the metric $g_{ij}$.

 In principle, it is enough to
compute $R_{ij}$ and $R_{ijkl}$ from the metric $g_{ij}$ to obtain these
recursion relations. In practice, these calculations are tedious
and some formal algebraic work has to be done first. The trick is to get rid
in the calculation of any dependence on the coordinates ${\pi ^i}$ by going
from the manifold itself to its tangent space, eq.(\ref{toto}). The crucial
advantage is
that in tangent space, the Riemann and Ricci tensors are functions  only of
the structure constants ${f_{ij}^k}$ of $ Lie(G)$ and that the tangent
space metric $\eta _{IJ}$ is constant, see eq.(\ref{eta}), and involves only
the coupling constants. In the vielbein basis,
eq.(\ref{evolution},\ref{beta}) becomes:

\begin{equation}
{\partial \eta_{IJ}\over \partial l}= \beta_{IJ} ,
\end{equation}
\begin{equation}
\beta _{IJ}(\eta)=   -\epsilon \eta_{IJ} + R_{IJ} +
{1\over2} T\ R_{IPQR\ }R_{JPQR} +
O(T^2).
\label{betaij}
\end{equation}
The matrix $\eta_{IJ}$ is given in eq.(\ref{eta}) and the  Riemann tensor
in tangent space can be expressed as:
\begin{eqnarray}
R_{IJKL} \hspace{-0.3cm}&=&\hspace{-0.3cm}f{_{IJ}}^af_{aKL} +
{1\over 2}{f_{IJ}}^M\left( f_{MKL}+f_{LMK}-f_{KLM}\right)\nonumber\\
\hspace{-0.3cm}& &\hspace{-0.3cm}+{1\over4}\left(f_{IKM} +f_{MIK}-f_{KMI}
\right)\left(
{{f_J}^M}_L + {f_{LJ}}^M -{f^M}_{LJ}\right)\nonumber\\
 \hspace{-0.3cm}& &\hspace{-0.3cm}-{1\over4}\left(f_{JKM} +f_{MJK}-f_{KMJ}
\right)\left(
{{f_I}^M}_L + {f_{LI}}^M -{f^M}_{LI}\right).
\label{riemann}
\end{eqnarray}
The indices $a$ and $\{I,J\dots\}$ refer to $H$ and  and $G-H$ respectively.
$G-H$ indices are raised and lowered by means of $\eta ^{IJ}$ and
$\eta _{IJ}$ and  repeated indices are summed over.
In NL$\sigma$ models the $\beta$ function and its derivatives allows
to compute the fixed point and the critical exponent $\nu$. Note that
since the $\beta$ function
is a tensor, it does not depend on a particular choice of coordinates.
As a consequence, the mere existence of a fixed point as well as the value
 of the exponent $\nu$ {\sl do not depend on the representation spanned
by the order parameter}.
 The other renormalization group function which is needed to give a
complete description
of the critical behavior is the Callan-Symanzik $\gamma$-function. This
function is determined by the field renormalization $Z$:
\begin{equation}
\gamma = -{\partial\log Z\over\partial l}.
\label{defgamma}
\end{equation}
 From this function follows the anomalous dimension $\eta$:
\begin{equation}
\eta=\gamma(\eta_1^{*},\eta_2^{*}) -\epsilon
\end{equation}
where $\eta_1^{*},\eta_2^{*}$ are the fixed point values of the
coupling constants.

\noindent The factor $Z$  is given at one loop order by the Laplace-Beltrami
operator acting on the coordinate $\pi^i$. It can be shown that this
 is nothing but $g^{ij}\Gamma_{ij}^k$ where $\Gamma_{ij}^k$ is the Christoffel
 connection on the metric manifold $G/H$:

 \begin{equation}
 Z\pi^k= \pi^k + {1\over \epsilon}g^{ij}\Gamma_{ij}^k .
\label{zz}
 \end{equation}
Once again, it is simpler to compute $g^{ij}\Gamma _{ij}^k$ by working in
tangent space. We find for any coset $G\otimes X/H\otimes X$ where $X$
is the subgroup of $G$ that commutes with $H$:
\begin{equation}
g^{ij}\Gamma_{ij}^k
=-{1\over\eta_1}\left(\sum_AT_AT_A\right)\pi^k+{\eta_2-\eta_1\over
\eta_1\eta_2}\left(\sum_\alpha T_\alpha T_\alpha\right)\pi^k ,
\label{ggamma}
\end{equation}
where $\{T_A\}$ and $\{T_\alpha \}$ are generators of $G$ and $X$ and
where $\eta_1, \eta_2$ are defined in equation (\ref{genac}). $\sum_AT_AT_A$
and $\sum_\alpha T_\alpha T_\alpha$ are Casimir operators of $G$ and $X$.
In general, a choice of coordinates is not stable under
renormalization. Equations (\ref{zz}) and (\ref{ggamma})  show that a good
coordinate system which renormalizes
 multiplicatively consists in the $\pi$ fields together with the
massive $\sigma$ modes. They build
up a linear representation of  $G\otimes X$ such that the $\pi$'s are an
eigenvector of the
 Casimir operators with an  eigenvalue that  depends on the representation.
{\sl Therefore, the $\gamma$-function and thus the critical exponent $\eta$
 depends on the representation $r$ of $G\otimes X$ spanned by the order
parameter}. In our case, the Casimir operators  have to be taken in the
vector representation of
both the $O(N)$ and the $O(2)$ groups. Their values are therefore
respectively $N-1$ and 1.
To summarize, in NL$\sigma$ models the existence of a fixed
point depends only on the symmetry breaking pattern
$G/H$ and not on the representation spanned by the order parameter. However,
the universality class is completely determined once the representation
$r$ of $G$ spanned by the observable is known. This scheme is completely
different from what happens in the $4-\epsilon$ expansion where even the
$\beta$ function, and thus the mere existence of a fixed point,
{\it does depend} on the representation of $G$ spanned by the order parameter.
In the following, we  apply these results to the $O(N)\otimes O(2)/O(N-2)
\otimes O(2)$ models. For reasons that will soon become clear, we shall
distinguish between the $N=3$ and $N>3$ cases.

\subsection{Results for $N>3$}

Using Eq.(\ref{betaij}) we obtain the following two loop recursion
relations valid for any $N\ge3$:
\begin{equation}
\left\{
\begin{array}{lll}
{\displaystyle{
{\partial\eta_1\over\partial l}}}& \hspace{-0.3cm}=-&\hspace{-0.3cm}\epsilon
 \eta_1 + N-2 -{\displaystyle{
\frac{1}{2}
\frac{\eta_2}{\eta_1} +
\frac{3N-4}{8}\frac{\eta_2^2}{\eta_1^3}+3(1-\frac{N}{2})\frac{\eta_2}
{\eta_1^2}}}\\
&&\\
 & & +{\displaystyle{(3N-8)\frac{1}{\eta_1}}}\\
&&\\
{\displaystyle{\frac{\partial\eta_2}{\partial l}}} & \hspace{-0.3cm}=
-&\hspace{-0.3cm}\epsilon
 \eta_2 + {\displaystyle{
\frac{N-2}{2}\left(\frac{\eta_2}{\eta_1}\right)^2  +\frac{N-2}{8}
\frac{\eta_2^3}{\eta_1^4}}}
\end{array}
\right.
\label{eqrecursion}
\end{equation}
Defining
$T_{1,2}=1/\eta_{1,2}$, we find that, apart from the trivial zero temperature
line of fixed points: $T_1=T_2=0$ with $T_1/T_2$ arbitrary
 there is one non trivial fixed point $C_{NL}$ with coordinates:
\begin{equation}
\left\{
\begin{array}{ll}
T_1^{*}\hspace{-0.3cm}&={\displaystyle{\frac{N-1}{ (N-2)^2}\left(\epsilon -
\frac{1}{  2}\frac{3N^2 -10N +4}{(N-2)^3}\epsilon^2\right) +  O(\epsilon^3)}}\\
&\\
T_2^{*}\hspace{-0.3cm}&={\displaystyle{\frac{1}{2}\frac{(N-1)^2}{ (N-2)^3}
\left(\epsilon - \frac{1}{2}\frac{5N^2 -16N +4}{(N-2)^3}\epsilon^2\right) +
O(\epsilon^3)}}
\label{pointfixe}
\end{array}
\right.
\end{equation}
This fixed point has one direction of instability so that  our model
undergoes an ordinary
second order phase transition with critical exponent $\nu$:
\begin{equation}
\nu^{-1}=\epsilon + {1\over2} {6N^3 - 27N^2 +32N-12\over (N-2)^3(2N-3)}\epsilon
^2+ O(\epsilon^3)
\label{nu}
\end{equation}
In order to complete our discussion, we have to specify the representation $r$
of $O(N)\otimes
O(2)$ spanned by the observable of the physical system under study. We are
interested in the
AFT model with $N$-component spins. In this case, the order parameter
transforms under the vector
representation of both $O(N)$ and $O(2)$, see Eq.(\ref{transphi}). At one loop,
it follows
 from Eq.(\ref{defgamma}) and  Eq.(\ref{ggamma}) that the anomalous dimension
$\eta$ is:

\begin{equation}
\eta ={3N^2-10N+9\over 2(N-2)^3}\epsilon +O(\epsilon^2)
\label{gamma}
\end{equation}

\subsection{Results for $N=3$}
Although both the recursion relations and the values of the exponents given in
the preceding
section are still valid in the $N=3$ case the symmetry properties are less
obvious. In this case,
we can take advantage of the different equivalent parametrizations of the
action we have derived
in section 2. The convenient parametrization is given by Eq.(\ref{actionp}):

\begin{equation}
{\cal S}_2={1\over2}\int d^Dx\  Tr\left(P(\partial R^{-1})(\partial R)\right)
,
\end{equation}
where $P$ is the diagonal matrix: $P=diag (p_1,p_2,p_3)$ and $R\in SO(3)$. In
the $O(3)\otimes O(2)/ O(2)$ case we have $p_1=p_2\ne p_3$. The relationship
between the $p_i$'s and the tangent space couplings $\eta_{IJ}$ is given in
Appendix B. Using Eq.(\ref{eqrecursion}) we can deduce
the  two loop recursion relations for the couplings $p_i$.
At the fixed point we find $p_1^{*}=p_2^{*}= p_3^{*}$ and thus $P^{*}\propto
1$. It follows from the discussion given in section 2 that the
 action ${\cal S}_2$ is
$O(3)\otimes O(3)$ symmetric at the fixed point: the symmetry has been
dynamically enlarged at
the fixed point. Since $O(3)\otimes O(3)/ O(3)\sim O(4)/O(3)$ the critical
behavior of the
 $O(3)\otimes O(2)/ O(2)$ NL$\sigma$ model is given by that of the $O(4)/O(3)$
 NL$\sigma$ model. It is a new result to find such a $O(4)$ symmetry for a
Heisenberg system. We
stress that it is not trivial to identify such a symmetry using a different
parametrization such as the one given in action
 ${\cal S}_3$ (see Eq.(\ref{actioncourant})). In
this case, the $O(4)$ symmetry is non-linearly realized on the fields ${\bf
e}_1,{\bf e}_2$.
The critical exponents  $\nu$ and  $\eta$ are given by
Eqs.(\ref{nu},\ref{gamma}) with $N=3$.
 Although the critical exponent $\nu$ is identical
to that of the $N=4$ vector model, it is not so simple to get $\eta$.
The order parameter is a  matrix
 $R(x)=({\bf e}_1(x),{\bf e}_2(x),{\bf e}_3(x)))$ (Eq.(\ref{actionp})) and
spans the {\it tensor} representation of $O(4)$.
This point was previously missed in ref.\cite{aza1}.
As a consequence, the exponent
$\eta$
 of the Heisenberg AFT model is the anomalous
dimension of a {\it composite} operator of the $N=4$ vector model. To see this,
 we need the relationship between the $O(3)$ matrix $R$  and a
 $O(4)$ unit vector. It can be shown that to any unit 4-component vector:
 \begin{equation}
\Psi =(\Psi_0, \Psi_i)\ \ ;\ \ \Psi_0^2+\sum_i \Psi_i^2=1
\end{equation}
there exists a matrix $R$ of $O(3)$ with components:
\begin{equation}
R_{ij}= 2(\Psi_i \Psi_j -{1\over4}\delta_{ij}) + 2\epsilon_{ijk}\Psi_0\Psi_k +
2(\Psi_0^2-{1\over4})\delta_{ij}
\end{equation}
Therefore, the expectation values of the vectors $<{\bf e}_i(x)>, i=1,3$
are obtained from those of the {\it bilinear} forms  $<(\Psi_i \Psi_j
-{1\over4}\delta_{ij})>$.

We thus find no new universality class for Heisenberg canted models but
instead the
general phenomenon of increased symmetry at the fixed point. These models
belong to the standard $N=4$ Wilson-Fisher universality class.
In dimension $D=3$, the exponent  $\nu$  is very accurately known
 \cite{leguillou}: $\nu=0.74$. However, the anomalous dimension
of the composite
operator $(\Psi_i \Psi_j -{1\over4}\delta_{ij})$ is only known at the two
loop order in $\epsilon=4-D$.
Let us finally emphasize that the phenomenon of increased symmetry at the
fixed point is particular to the $N=3$ case in the
$O(N)\otimes O(2)/O(N-2)\otimes O(2)$ NL$\sigma$ models. For any $N>3$,
the phase transition belongs indeed
 to a universality class different from $O(N)$
but as one reaches the physical $N=3$ case one falls in the
well known $O(4)$ one. This conclude our analysis of the NL$\sigma$
models associated to canted magnets.

The well-known $\epsilon = 4-D$ expansion
starting from the upper critical dimension
of the appropriate  Landau-Ginzburg-Wilson (LGW) action
has been  applied to helimagnets more than ten years
ago by Garel and Pfeuty\cite{garel} and Bailin et al.\cite{bailin}. More
recently, renewed interest
on this subject has been drawn by Kawamura\cite{kawamura}.
We shall, in the next section, present the results obtained from this
expansion.

\section{The linear theory and the $\epsilon = 4-D$ expansion}

The LGW action can be obtained in the same spirit as the NL$\sigma$. Once the
symmetry breaking pattern is known, in our case $O(N)\otimes O(2)/O(N-2)\otimes
O(2)$,
all  we have to do is to find the most general action which is $O(N)\otimes
O(2)$
symmetric and which  ground state is $O(N-2)\otimes O(2)$ invariant. Among all
possible actions, one has to select those which are
{\it renormalizable} in $D=4$, i.e. to keep only terms up to order 4
 in the fields and to order 2 in their derivatives.
The LGW action  does not possess the invariance under
reparametrization of the  NL$\sigma$ model. Moreover, only linear
transformations of  $O(N)\otimes O(2)$ are allowed since non-linear
transformations involve higher powers of the fields and their derivatives
than allowed by
renormalizability. For this reason the whole LGW or Linear
 theory depends explicitly on the representation of $O(N)\otimes O(2)$ spanned
by the
physical order parameter. This fact  have dramatic consequences on the
renormalizability of LGW theories as compared to their corresponding NL$\sigma$
models.
In order to build the LGW action, we shall start from the
NL$\sigma$ model. In the particular case of canted models,
 we have to choose the parametrization  which spans a
linear representation of  $O(N)\otimes O(2)$. In this case the partition
function Z
is given by:

\begin{equation}
Z=\int D{\bf e }_1 D{\bf e }_2\  \delta( {\bf e }_1.{\bf e}_2)\ \delta(
{{\bf e }_1}^2 -1)\
\delta( {{\bf e }_2}^2 -1)e^{-{\cal S}_3},
\label{actioncourantbis}
\end{equation}
\begin{equation}
{\cal S}_3 ={1\over2}\int d^Dx \left(g_1\left(\nabla
{\bf e}_1^2 + \nabla
{\bf e}_2^2\right)
+ g_2\left({\bf e}_1\nabla{\bf e}_2 - {\bf e}_2\nabla{\bf e}_1\right)^2\right)
{}.
\end{equation}
The ${\bf e}_i$ are $N$-component vectors. The LGW action is now obtained in
a standard way by relaxing the constraints in Eq.(\ref{actioncourantbis})
and use of a potential:
\begin{equation}
V({\bf e }_1,{\bf e }_2)=
{1\over2} m^2 ({\bf e }_1^2+{\bf e }_2^2) +
u_1 ({\bf e }_1^2+{\bf e }_2^2)^2 + u_2 ({\bf e }_1\wedge{\bf e }_2)^2
\end{equation}
The LGW action for canted magnets reads now:

\begin{equation}
{\cal S}_{LGW} ={1\over2}\int d^Dx \left({1\over2}\left(\nabla
{\bf e}_1^2 + \nabla
{\bf e}_2^2\right)
+ V({\bf e }_1,{\bf e }_2)\right)
\label{actionlin}
\end{equation}
We have rescaled the fields in order to obtain the standard normalization for
the gradient
term and have omitted the current term
$({\bf e}_1\nabla{\bf e}_2 - {\bf e}_2\nabla{\bf e}_1)^2$
 since it is {\it not}
renormalizable. Note that it is precisely this term which allowed the
NL$\sigma$
action ${\cal S}_3$  to be $O(3)\otimes O(3)/O(3)$ symmetric at the fixed point
when
$N=3$.

The two loop recursion relations for the couplings $u_1$ and $u_2$ were first
obtained by Bailin et al.\cite{bailin} and Garel and Pfeuty\cite{garel} . We
shall here only summarize their results.
Let us comment the RG flow:

i) there is a critical value of $N$ depending on the dimension:
$N_c(D)=21.8-23.4\epsilon+ O(\epsilon^2)$, under which there is no fixed point.
In this
case the transition is expected to be first order. Let us emphasize that for
$\epsilon =1$ the second term in the $\epsilon$-expansion of
$N_c(\epsilon)$ is {\it not} a small perturbation of the first one since it is
$-23.4$. This is the signal that a precise determination of $N_c(D)$ needs
some control of the $\epsilon$-expansion which, as it stands, can not be
used directly for $\epsilon$ of order 1.

ii) for $N>N_c(D)$ there are still two different regions in the portion of
the $(u_1 ,u_2)$
parameter space where the potential is stable: see fig. 1.
One is
the second order region. It lies above the line $L$ joining the origin to
an unstable fixed point (called
$C_-$) and is the basin of attraction of the stable fixed
point, called
$C_L$. The Heisenberg fixed point $H$ is unstable towards $C_L$.
The other region lies between the line
$L$ and the stability line $S$ of the potential: $u_2=-2u_1$. It is a region
of runaway behaviour
 and is expected to correspond to a first order region. We note
that, as $N$ tends to infinity, $L$ tends to $S$, as expected.

\section{Interpolating between $D=2+\epsilon$ and $D=4-\epsilon$.}

There is clearly a mismatch between RG results obtained in
$D=4-\epsilon$ dimensions from the LGW model and in $D=2+\epsilon$
dimensions from the $O(N)\otimes O(2)/O(N-2)\otimes O(2)$ sigma model.
Even though these NL$\sigma$  models predict for any $N$ a continuous
transition in the neighborhood of dimension 2, there are models with
either $N<N_c$ or which do not belong to the basin of attraction of $C_L$
for which the transition is expected to be of first order at least near
$D=4$.  This is very different from the ferromagnetic case where both the
$O(N)/O(N-1)$ NL$\sigma$ model and the LGW model predict the same
critical behaviour. However, when $N>N_c(D)$, there exists a domain in the
coupling constants space $M$ where a second order transition is predicted
in both models. These domains are respectively the basins of
attraction of $C_{NL}$ in $D=2+\epsilon$ and of $C_L$ in $D=4-\epsilon$.
The natural question is whether or not these two fixed points are the same
in a given dimension $D$ between $2$ and $4$. The $1/N$ expansion allows
to answer this question, at least for $N$ large enough, since this expansion is
non
perturbative in the dimension \cite{aza2}. The critical exponent $\nu$ has been
calculated to the lowest non trivial order in a $1/N$ expansion of the
Landau-Ginzburg
action (\ref{actionlin})\cite{bailin,kawamura} :
\begin{equation}
\nu_{1/N}(D) = {1\over D-2}\left(1-{1\over ND}12(D-1)S_D\right),
\label{nu1/N}
\end{equation}
\begin{equation}
S_D={\sin \left({\pi\over 2}(D-2)\right)\Gamma(D-1)\over 2\pi\left(
\Gamma(D/2)\right)^2}.
\end{equation}
By expanding eq.(\ref{nu1/N}) to lowest non trivial order in $\epsilon$,
$\epsilon =4-D$ or $\epsilon =D-2$, we find that $\nu _{1/N}(D)$ coincides
with $\nu_{4-\epsilon}(N)$ and $\nu_{2+\epsilon}(N)$ to lowest order in
$1/N$. The same type of expansion can be done on the other exponents with the
same results.

We may thus conclude as in the ferromagnetic case that, when the fixed point
exists near $D=2$ and
near $D=4$, we can follow it smoothly from $D=4-\epsilon$ down to
$D=2+\epsilon$. Therefore, in the
whole space $E=$\{$(M)$ = coupling constants, $D, N$) there should exist a
domain $Z$ where the
transition is of second order and which is governed by a unique fixed
point $C_L(N,D)=C_{NL}(N,D)$.
In the complementary of $Z$ the transition is expected to be of
first order. On the boundary
$\Gamma$ of these two domains, the transition should be tricritical
in the simplest hypothesis.

The situation can be summarized in the plane (N, D) of number
of components of the model and dimension.
The $4-\epsilon$ findings have shown that there is a universal
curve $N_c (D)$ separating a first-order region and a second-order
region. If one believes that the $2+\epsilon$ results survive
perturbation theory then the neighborhood of $D=2$ belongs
to the second-order region for all $N\ge 3$. As a consequence,
the line $N_c (D)$ intersects the $N=3$ axis somewhere
between $D=2$ and $D=4$. This defines thus a critical dimension $D_c$
that we do not expect to be a simple number.
Making the hypothesis that the RG calculations have captured all the
relevant fixed points then
there are two possibilities:

\noindent
i) The critical value is between $D=3$ and $D=4$. This implies
that the physical case $N=3$, $D=3$ undergoes an O(4) transition
as shown in (4.3). This case is favored by present numerical studies
\cite{bata}.

\noindent
ii) The critical value is between $D=2$ and $D=3$. Then the physical case
is governed by a fluctuation-induced first-order transition.
It cannot be excluded that $D_c =2$ in which case the perturbative
analysis of the nonlinear sigma model is always irrelevant.

We note in addition that there is an additional possibility namely
$D_c =3$. We do not see any reason why this would be realized
since this looks like an artificial fine-tuning. In this case
one may speculate that a tricritical mean-field like behaviour
is seen for $N=3$, $D=3$.

These alternatives are consistent with all the RG results and
do not require additional fixed points not seen in perturbation theory.
In this picture the stable fixed point seen in $4-\epsilon$
for $N\ge N_c$ can be followed smoothly by the large-N limit
till $D=2$ and then identified with the conventional O(4) fixed point
via the $2+\epsilon$ calculation of (4.2-3) for $N=3$.

\section{Conclusion}
We have  shown in this article that the non linear sigma model provides a new
approach to the analysis of the critical behavior of frustrated systems.
The double expansion in $T$ and in $\epsilon=D-2$ of the $O(N)\otimes
O(2) /O(N-2)\otimes O(2)$ NL$\sigma$ model
has been performed and a fixed point has been
found in $D=2+\epsilon$ for any $N$, which turns out to have a remarkable
$O(4)$ symmetry for three component spins. Since for $N\le 21$ the transition
is expected to be first order near $D=4$, we conjecture that for any $N\in
[3,21]$, the nature of the phase transition changes between 2 and 4 dimensions
and is tricritical at the border of the second and first order region.

In the simplest hypothesis the (N, D)-plane is divided
in two region: a first-order region containing $N=3$ and $D=4$
and a second-order region containing the $D=2$ line (for any N),
the whole large-N line (for all D) and also in the neighborhood
of D=4 the $N\ge 21.8$ points. In between lies a universal line
$N_c (D)$ whose $4-\epsilon$ expression was already known.
This universal line intersects the N=3 axis for some unknown
critical dimension $D_c$.
If $3<D_c <4$ then the physical point N=3, D=3 is second order
in the O(4) universality class and its exponent $\eta$ is that
of a {\it tensor} representation.
If $2 < D_c < 3$ the physical point is first-order.
It may happen that $D_c =3$ (although we see no reason why)
in which case one could see tricritical behaviour.
This phase diagram is in agreement with all known RG results.
To decide the fate of the physical point requires clearly
additional techniques. Present Monte-Carlo results\cite{bata}
favor $2 < D_c < 3$ although more work is needed.
Direct RG calculations in D=3 may also help to confirm the phase
diagram\cite{aza3}.

We note that in the generic case
$O(N)\otimes O(2) /O(N-2)\otimes O(2)$, $N>3$, the symmetry
is {\it not}  enlarged at the fixed point.
In this respect $N=3$ is exceptional: for other values of N,
the fixed point does not belong to the O(N) Wilson-Fisher family.
It would be interesting of course to investigate the fate of the
XY N=2 case since known helimagnets have significant anisotropies
that lead to a non-Heisenberg behaviour.

\bigskip
\bigskip

\noindent
{\bf Acknowledgements:}

\bigskip

Two of us (P. A. and B. D.) would like to thank A. Caill\'e,
H. T. Diep and M. L. Plumer for numerous discussions.
One of us (Th. J.) would like to thank Th. Garel, J. C. Le Guillou and A. P.
Young for several discussions and also D. P. Belanger
for informing us about the present experimental status of
phase transitions helimagnetism.
\newpage
{\bf APPENDIX  A}

We have seen in section 6 that for $N=3$ the LGW  and  NL$\sigma$ models
do not predict the
same critical behaviour. No fixed point is found in $D=4-\epsilon$
dimensions from the LGW action
while the NL$\sigma$ model predicts a fixed point in $D=2+\epsilon$ with a
$O(3)\otimes O(3)\sim O(4)$ symmetry at this point. Actually, though
these results are
perturbative, it is easy to see that a $O(4)$-symmetric fixed
point {\sl can not be obtained}
with the LGW action (\ref{actionlin}) since no value of $(u_1,u_2)$ makes
this action
$O(4)$-symmetric. The reason is that the rectangular
matrix $(\phi_1,\phi_2)$ represents 6
degrees of freedom on which acts $O(3)$ on the right and $O(2)$ on the
left and that this
$O(2)$ can not be enlarged to $O(3)$ with only these 2 fields. This $O(3)$
symmetry can be
realized only with at least 3 fields: $(\phi_1,\phi_2,\phi_3)$.

This would correspond to  the 9-dimensional representation of $O(4)$. To
check whether the
discrepancy between the two models can be eliminated by allowing the
LGW model to reach the
$O(3)\otimes O(3)$ symmetry, we have built and studied the most general
action invariant under
$O(3)\otimes O(2)$ and involving 3 fields:
\begin{equation}
\begin{array}{ll}
H=  &\hspace{-0.3cm}{1\over2}(\partial\phi_1)^2 + {1\over2}(\partial\phi_2)^2 +
 {1\over2}(\partial\phi_3)^2 + {1\over4}u_1\left( \phi_1^2 + \phi_2^2\right)^2+
{1\over2}u_2\left( \phi_1^2  \phi_2^2 - (\phi_1.\phi_2)^2\right)\\
&\\  &\hspace {-0.3cm}+ {1\over4}u'(\phi_3^2)^2+ {1\over2}u_4\phi_3^2
\left( \phi_1^2 +
 \phi_2^2\right) - {1\over2}u_3\left( (\phi_1.\phi_3)^2 +
(\phi_2.\phi_3)^2\right)
\end{array}
\label{action3champs}
\end{equation}
$(\phi_1,\phi_2)$ is a doublet of $O(2)$ and $\phi_3$ a singlet.
$H$ is $O(3)\otimes O(3) $ invariant when :
\begin{equation}
u_2=u_3\ \ ;\ \ u_1+u_2=u_4\ \ ;\ \ u_1=u'
\label{conditions}
\end{equation}
A one-loop RG calculation shows that no attractive fixed point exists in
$D=4-\epsilon$ in this model. This is the proof that the root of the problem is
not only a question of symmetry but also a problem of dimension,  field
content and renormalizability. More precisely, if we set  $u_2,u_4$ and $u'$ to
the values eq.(\ref{conditions}) which make hamiltonian
eq.(\ref{action3champs})
$O(3)\otimes O(3)$-symmetric, we find that the remaining symmetry in the broken
phase is $O(3)_{diag}$. Then, the NL$\sigma$ model associated with this
symmetry
breaking scheme is $O(4)/ O(3)$. This NL$\sigma$ model is unique since it
depends only on the Goldstone modes, and then only on the Lie algebras of
$O(4)$ and $O(3)$ and {\it not} on the representations of these groups. On the
other hand, there are as many associated LGW models as there are
representations
of $O(4)$ (or at least as there are actions built with representations
of $O(4)$
that can be broken down to $O(3)$). Surprisingly, the LGW action built with the
4-component vector representation of $O(4)$ admits a fixed point in
$4-\epsilon$
dimensions (the usual Heisenberg fixed point) and that built with the
9-dimensional tensor representation $(\phi_1,\phi_2,\phi_3)$ admits no such
fixed point, as we have seen
above. Since in these two LGW models, the symmetry breaking scheme is the same
and then the Goldstone modes are the same, it means that the difference between
these models lies in the massive modes. It is not clear up to now whether these
modes can indeed be physically relevant for the critical behaviour. At least
perturbatively and near 2 dimensions, the NL$\sigma$ model does not take care
of these modes.  In our case, we have to deal with the vector and tensor
representations of $O(4)$ which both allow to replace the constraints of the
NL$\sigma$  model by potentials and which lead to two different results. The
relevance of the massive modes in the critical behaviour is then directly
related to the way one chooses to go from the microscopic Hamiltonian to the
different continuous actions, linear or non linear.

\noindent
{\bf APPENDIX B}

We give in this appendix the relationship between different
parametrizations of the sigma model.

\noindent
{\bf From tangent space to the constraints (for any $N$):}

   We parametrize the matrix $L$ in Eq.(\ref{LpiO}) as:
    \begin{equation}
    L = (A\ \phi_1\ \phi_2)
    \end{equation}
    where $A$ is a rectangular $N\times (N-2)$ matrix, $\phi_1$
and $\phi_2$ are
    $N$-component vectors. Since $L$ is in $O(N)$, they must satisfy:
    \begin{equation}
    ^tA.A=1_{N-2}\ \ ,\ \ ^tA\phi_1=^tA\phi_2=0\ \ ,\ \ \phi_1^2=
\phi_2^2=1\ \ ,\ \   \phi_1 .\phi_2=0
    \end{equation}
    \begin{equation}
    A.^tA + \phi_1.^t\phi_1 + \phi_2.^t\phi_2 = 1_N
    \end{equation}
We are interested in the action of the NL$\sigma$ model so that we have
to compute $ L^{-1}\partial L$:
    \begin{equation}
    L^{-1}\partial L=
    \left(
    \begin{array}{ccc}
    0& ^tA\partial\phi_1&^tA\partial\phi_2\\
    -^tA\partial\phi_1& 0&\phi_1.\partial\phi_2\\
    ^tA\partial\phi_2&-\phi_1.\partial\phi_2&0
    \end{array}
    \right)
    \end{equation}
    The projection of $L^{-1}\partial L$ onto $Lie(G')-Lie(H')$
leads to two
sets of
    vielbein that are not mixed under the H-transformations:
     \begin{equation}
     (L^{-1}\partial L)_{\vert_{G'-H'-X}}=
    \left(
    \begin{array}{ccc}
    0& ^tA\partial\phi_1&^tA\partial\phi_2\\
    -^tA\partial\phi_1& 0&0\\
    ^tA\partial\phi_2&0&0
    \end{array}
    \right)
    \end{equation}
    and
     \begin{equation}
     (L^{-1}\partial L)_{\vert_{X}}=
    \left(
    \begin{array}{ccc}
    0& 0&0\\
    0& 0&\phi_1.\partial\phi_2\\
    0&-\phi_1.\partial\phi_2&0
    \end{array}
    \right)
    \end{equation}
  The total action is the sum of the traces of the squares of these
    two matrices weighted by $\eta_1$ and $\eta_2$. These
    coefficients are by definition those  coming from the tangent
    space metric $\eta_{IJ}$.
    \begin{equation}
    \begin{array}{l}
    \eta_1 Tr\left[\left((L^{-1}\partial L)_{\vert_{G'-H'-X}}\right)^2\right]
    =-2\eta_1\left((\partial\phi_1)^2 +(\partial\phi_2)^2\right) +
    4\eta_1(\phi_1.\partial\phi_2)^2\\
    \eta_2 Tr\left[\left((L^{-1}\partial L)_{\vert_{X}}\right)^2\right]
    =-2\eta_2(\phi_1.\partial\phi_2)^2
    \end{array}
    \end{equation}
   The coupling constants $\eta_1$, $\eta_2$ are now easily related to
$g_1$ , $g_2$ defined in (\ref{actioncourant}):
\begin{equation}
    \left\{
    \begin{array}{ll}
    \eta_1 &={g_1/2}\\
    \eta_2&= g_1 +2g_2
    \end{array}
    \right.
    \end{equation}

\noindent
{\bf From the P-matrix to the constraints ($N=3$)}

    We now compute directly $Tr (P(R^{-1}\partial R)^2)$ to obtain
    explicitly the action (\ref{actionp}). Let us start with $R\ \in\  O(3)$,
    parametrized as follows:
    \begin{equation}R = (e_1\ e_2\ e_1\wedge e_2)
    \label{r}
    \end{equation}
    where $e_1$ and $e_2$ are two 3-component vectors such that
 $e_1^2=e_2^2=1$ and $e_1 .e_2=0$. The diagonal part of $(R^{-1}\partial
    R)^2$ is:
    \begin{equation}
    (R^{-1}\partial R)^2_{\vert_{diag}}=\left(
    \begin{array}{lll}
    -(\partial e_1)^2& & \\
     &-(\partial e_2)^2& \\
     & &-(\partial e_1)^2-(\partial e_1)^2 +2(\partial e_1.e_2)^2
    \end{array}
    \right)
    \end{equation}
    so that:
    \begin{equation}
    Tr\left(P(R^{-1}\partial R)^2\right) = -(p_1+p_2)\left((\partial e_1)^2
+ (\partial e_2)^2\right)
 +{p_2\over2}(\partial e_1.e_2- \partial e_2.e_1)^2
    \end{equation}
 We obtain the relations between $(p_1,p_2)$ and $(g_1,g_2)$:
    \begin{equation}
    \left\{
    \begin{array}{ll}
    p_1 &= g_1 +2g_2 ,\\
    p_2 &= -2g_2 .
    \end{array}
    \right.
    \end{equation}

\newpage
\bibliographystyle{unsrt}
\bibliography{heli}
\newpage
\centerline{\bf FIGURE CAPTIONS}
\bigskip
\bigskip
\noindent
{\bf Figure 1:}

\noindent
The renormalization group flow in the neighborhood of four
dimensions for the Landau-Ginzburg model (\ref{actionlin}) in the
case $N\ge N_C (D)$.
On the $u_1$ axis one finds the conventional O(2N) Wilson-Fisher
fixed point which is unstable towards the fixed point $C_L$.
By following a smooth path in the (N, D) plane we find that
$C_L$ is the O(4) fixed point when N=3 in the neighborhood of D=2.
\end{document}